\documentclass[a4paper,fleqn,usenatbib]{mnras}

\usepackage{mathptmx}

\usepackage[T1]{fontenc}
\usepackage{ae,aecompl}

\usepackage{graphicx}	
\usepackage{amsmath}	
\usepackage{amssymb}	
\usepackage{longtable}

\newcommand{\ha}{H$\alpha$}

\newcommand \vhel{\ifmmode{~V_{{\rm HEL}}}\else{~$V_{{\rm HEL}}$}\fi}

\title[NGC~6778: Extreme adfs and CSPN binarity]{NGC~6778: Strengthening the link between extreme abundance discrepancy factors and central star binarity in planetary nebulae}

\author[D. Jones et. al]{D. Jones,$^{1, 2}$\thanks{E-mail:
djones@iac.es} R. Wesson,$^{3}$ J. Garc\'ia-Rojas,$^{1, 2}$ R.L.M. Corradi, $^{1, 2}$ and 
\newauthor H.M.J. Boffin$^{3}$    
\\
$^{1}$Instituto de Astrof\'isica de Canarias, E-38205 La Laguna, Tenerife, Spain\\
$^{2}$Departamento de Astrof\'isica, Universidad de La Laguna, E-38206 La Laguna, Tenerife, Spain\\                       
$^{3}$European Southern Observatory, Alonso de C\'ordova 3107, Casilla 19001, Santiago, Chile\\
}

\date{Accepted 2015 October 26.  Received 2015 October 26; in original form 2015 October 7}

\pubyear{2015}

\begin{document}
\label{firstpage}
\pagerange{\pageref{firstpage}--\pageref{lastpage}}
\maketitle

\begin{abstract}
We present new optical spectra of the nearby, bright, planetary nebula NGC~6778.  The nebula has been known to emit strong recombination lines for more than 40 years but this is the first detailed study of its abundances.  Heavy element abundances derived from recombination lines are found to exceed those from collisionally excited lines by a factor of $\sim$20 in an integrated spectrum of the nebula, which is among the largest known abundance discrepancy factors.  Spatial analysis of the spectra shows that the abundance discrepancy factor is strongly, centrally peaked, reaching $\sim$40 close to the central star.  The central star of NGC~6778 is known to be a short period binary, further strengthening the link between high nebular abundance discrepancy factors and central star binarity.
\end{abstract}

\begin{keywords}
planetary nebulae: individual: NGC~6778 -- circumstellar matter -- stars: mass-loss -- stars: winds, outflows -- binaries: close -- ISM: abundances
\end{keywords}



\section{Introduction}

Abundance analyses based on deep spectroscopic observations of planetary nebulae (PNe), deep enough to measure the relatively weak optical recombination lines (ORLs) from heavy element ions, have been found to yield systematically higher ionic abundances when calculated from ORL fluxes than when calculated using the brighter forbidden lines (collisionally excited lines, CELs) of the same species.  In general, the ratio of ORL to CEL abundances, commonly known as abundance discrepancy factors (\emph{adf}s), are of order 1--5, with a very small number of PNe being found to exhibit \emph{adf}s one or two orders of magnitude greater \citep{liu06b,liu06,wesson05}.  The origin of these systematic discrepancies remains an unresolved problem in nebular astrophysics \citep[also appearing in studies of H~\textsc{ii} regions,][]{garcia-rojas07}, which has clear repercussions for the study of the chemical composition and evolution of the Universe \citep{carigi05,esteban14}.  It has been suggested that the highest \emph{adf}s may be due to a multiphase composition whereby the nebula consists of a phase of typical nebular composition and temperature ($\sim$10$^4$ K), where CELs can be excited efficiently, and a second phase of metal-enriched, lower temperature ($\sim$10$^3$ K) gas, where the ORLs form \citep[see e.g.\ ][]{liu00,liu06}.

Recently, \citet{corradi15} studied the abundances of three bright northern PNe which are also host to a close binary central star (Abell~46, Abell~63 and Ou~5).  Their analysis revealed that all three show \emph{adf}s greater than the average, with those of Abell~46 and Ou~5 being particularly extreme (reaching up to 300 in the inner regions of Abell~46).  Combining their findings with the studies of \citet[Hf~2-2]{liu06} and \citet[the proposed binary, Abell~30]{wesson03}, they conclude that of the most extreme PN \emph{adf}s are all found in PN which host close binary central stars.  However, the sample on which this claim is based is rather limited, with only a handful of PNe being shown to display \emph{adf}s greater than $\sim10-15$ (and only three of these being confirmed to host binary central stars\footnote{Abell~30 is strongly believed to host a binary central star, but no direct evidence has yet been found \citep{wesson03}.}).  Adding further objects to this sample is critically important for strengthening the link between binarity and extreme \emph{adf}s, and eventually enabling a quantitative analysis of the relationship between the two.

\begin{figure}
\centering
\includegraphics[width=5.5cm, angle = 90]{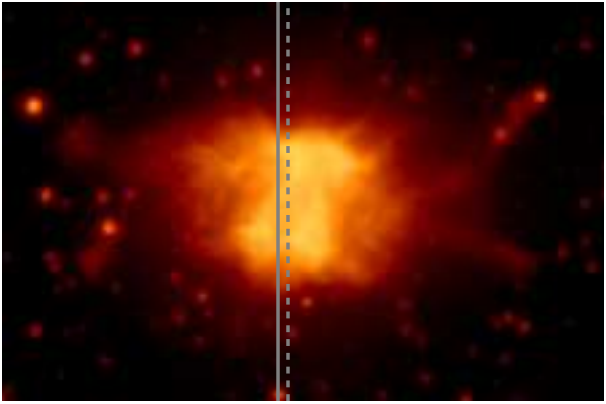}
\caption{VLT-FORS2 acquisition image of NGC~6778 in the light of \ha+[N~\textsc{ii}].  The image measures roughly 1.5\arcmin{}$\times$1\arcmin, the vertical axis represents a position angle of $-$55\degr{} with North through East in an anti-clockwise direction.  The positions of the blue (dashed line) and red (solid line) FORS2 longslits are overlaid, showing the approximately 1\arcsec{} offset between the two slit centres.}
\label{fig:image}
\end{figure}

\begin{figure*}
\centering
\includegraphics[angle=270,width=\textwidth]{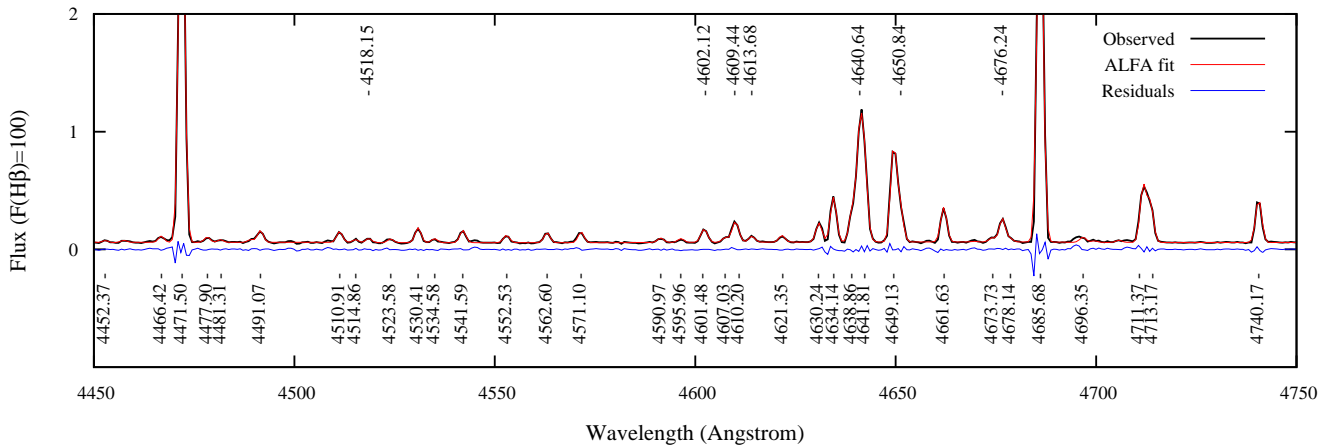}
\caption{A section of the observed VLT-FORS2 spectrum of NGC~6778 with the wavelengths of several of the detected lines highlighted (see table \ref{tab:fluxes} for the full list).  The observed spectrum is shown in black, while the \textsc{alfa} fit is shown in red and the fit's residuals in blue.}
\label{fig:alfafit}
\end{figure*}

NGC~6778 ($\alpha = 19^h 18^m 24.94^s$, $\delta = -01^\circ 35' 47.641''$, PN G034.5$-$06.7, Fig.~\ref{fig:image}) was found to show unusually strong ORLs, typical of high \emph{adf}s, in the study of \citet{czyzak73}, and was later shown to host a close binary central star with a period of only 3.7 hours \citep[making it one of the shortest period binary central stars known; ][]{miszalski11b}.  In this paper, we present a detailed abundance analysis of NGC~6778, revealing its \emph{adf} to be amongst the highest known, strengthening the link between extreme \emph{adf}s and central star binarity. The paper is structured as follows: Section~\ref{sec:obs} presents the observations and data reduction, Section \ref{sec:analysis} presents the methodology and results of the abundance analysis, and, finally, Section \ref{sec:discussion} contains a discussion of the results and conclusions drawn.

\section{Observations and data reduction}
\label{sec:obs}

NGC~6778 was observed on the night of June 18 2014 with the FORS2 instrument mounted on the ESO VLT's UT1 Antu telescope \citep{FORS}.  The instrument set-up consisted of a longslit measuring 0.7\arcsec $\times$6.8\arcmin{} at a position angle of $-$55\degr{}, and a mosaic of two 4k$\times$2k MIT/LL CCDs binned 2$\times$2 ($\equiv$0.25\arcsec{} per pixel).  A single 600s exposure was taken using the GRIS\_1200B grism, followed by a 60s exposure with the GRIS\_1200R grism and GG435 filter, offering an average resolution of 1.5\AA{} across a wavelength range of $\sim$3600--5000\AA{} and $\sim$5800--7200\AA{}, for the two grisms respectively.  The object was reacquired between exposures  resulting in a small offset of $\sim$1\arcsec{} between the spectra (see Fig. \ref{fig:image}); the seeing was approximately 0.9\arcsec{} for the duration of the observations.

The two-dimensional spectra were bias subtracted, cleaned of cosmic ray events, wavelength calibrated and flux calibrated against exposures of the standard star LTT7987 \citep
{hamuy92}, all using standard \textsc{starlink} routines \citep{ccdpack,figaro}.  The spectra were then sky subtracted and extracted to one-dimensional spectra using ESO-\textsc{midas}.  

We derived emission line fluxes using a new code, {\sc alfa} (Wesson, submitted), which optimises the parameters of Gaussian fits to line profiles using a genetic algorithm.  It fits all lines in the spectrum simultaneously after subtracting a globally-fitted continuum.  Fig.~\ref{fig:alfafit} shows the results of the fitting in the region of 4600--4700{\AA} where there are many recombination lines, and clearly demonstrates the effectiveness of \textsc{alfa}.  Table~\ref{tab:fluxes} lists the fluxes (both observed and dereddened) measured for all emission lines in the spectrum, together with their 1$\sigma$ uncertainties.  Note that, given the small spatial offset between red and blue exposures, a correction of 10\% was applied to the fluxes from the red exposure in order that the reddening as calculated from the H$\alpha$/H$\beta$ ratio matched that calculated from the H$\gamma$/H$\beta$ and H$\delta$/H$\beta$ ratios, which gave very similar values of c(H$\beta$). The possible uncertainties from this rescaling as well as the effect of the spatial offset on the results will be discussed further later.  Furthermore, it is important to note that the uncertainties in line fluxes determined using {\sc alfa} \emph{do not} include any contribution from systematic uncertainties, such as those originating from the flux calibration, but are purely statistical in nature (being based on the RMS of the residuals after subtracting the fitted lines).  As such, all uncertainties can only be considered lower limits given that in many, if not all regions of the spectrum, the uncertainty will be dominated by uncertainties in the relative flux calibration.

\section{Abundances and physical properties}
\label{sec:analysis}

To measure abundances from the emission line fluxes, we used the {\sc neat} code \citep{wesson12}, which employs a Monte Carlo technique to robustly propagate uncertainties from line flux measurements into chemical abundances.  The code corrects for interstellar extinction using the ratios of H$\gamma$ and H$\delta$ to H$\beta$ (the H$\alpha$/H$\beta$ was not used to calculate the extinction but rather to re-scale the red exposure to provide a similar measure of the extinction, see Section \ref{sec:obs}) and the Galactic extinction law of \citet{howarth83}; temperatures and densities are then derived from the standard diagnostics \citep[see][for full details]{wesson12}.  Elemental abundances are then calculated from flux-weighted averages of the emission lines of each species (using the previously derived temperatures and densities). Total chemical abundances were calculated using the ionisation correction scheme of \citet{delgado-inglada14}, except in the case of argon where the Ionisation Correction Function (ICF) from \cite{kingsburgh94} was used\footnote{The ICF from \citet{kingsburgh94} was adopted for argon as our spectrum shows no [Ar~\textsc{iii}] lines but does display [Ar~\textsc{iv}] emission.  The \citet{delgado-inglada14} scheme does not have an ICF for this case whereas the \citet{kingsburgh94} scheme does.}. Note that, as is common among similar studies, the uncertainties of these ionisation correction functions are not propagated in our analysis.
 
The physical parameters determined for NGC~6778 are listed in Table~\ref{tab:etd}, while the final ionic and total abundances, are listed in Tables~\ref{tab:ionic} and \ref{tab:abund}, respectively.  Reddening values in the literature range from $c$(H$_\beta$) $\sim$ 0.3 \citep{czyzak73,aller79, voigt65} through to $c$(H$_\beta$) $\sim$ 0.9 \citep{kaler70,cahn92}, however many of these values are based only on a single ratio taken with non-CCD instrumentation.  Our measured reddening, $c$(H$_\beta$) = 0.46, lies close to the centre of the literature range, and given that it is based on Balmer line ratios which give very consistent results (see above), and data of such high signal-to-noise, we infer that our value is the most reliable to-date.  Abundances are calculated using the temperature derived from [S~\textsc{ii}] for singly ionised species, and [O~\textsc{iii}] for more highly ionised species.  The [S~\textsc{ii}] temperature is derived using line pairs which are widely separated in wavelength and were thus observed with different grisms; the blue line pair at 4068/76{\AA} is also blended with O~{\sc ii} recombination lines from the V10 multiplet.  The statistical uncertainty derived by \textsc{neat} ($\sim$870 K) is thus likely to underestimate the true uncertainty.  However, in spite of its underestimated uncertainty, the [S~\textsc{ii}] temperature is found to agree well with that determined using [O~\textsc{iii}] (see below for further discussion on the uncertainties arising from the atomic data used).  The temperature determined from the Balmer Jump is much lower, $T_e=4090$ K, as is common in high \emph{adf} nebulae \citep{corradi15}.  This is generally considered to be evidence of the dual component (hot and cold) nature of the nebular plasma \citep[][further supported by the low temperature derived from the O~\textsc{ii} ORLs, which will be discussed further in section \ref{sec:hetemp}]{liu00}.


\begin{table*}
\caption{Observed, $F(\lambda)$, and dereddened, $I(\lambda)$, nebular emission line fluxes relative to H$\beta = 100$. Note that an upwards correction of 10\% has been applied to the fluxes from the ``red'' spectrum such that the reddening as derived from all Balmer line ratios is consistent (see text for further details).  Asterisks denote that the line is detected as a blend with the previous line in the table, and the flux listed is the total flux of the unresolved lines. The upper and lower term columns show the upper and lower levels of the atomic transition producing the line, while g$_1$ and g$_2$ are the degeneracy factors of the two levels.}
\label{tab:fluxes}
\begin{tabular}{lrlrlllllll}
\hline
    $ \lambda $ (\AA{}) & \multicolumn{2}{c}{$F \left( \lambda \right) $ }& \multicolumn{2}{c}{$I \left( \lambda \right) $} & Ion & Multiplet & Lower term & Upper term & g$_1$ & g$_2$\\
\hline
 3645.50 &   0.69& $\pm$   0.06&  0.93& $\pm$   0.08 & Balmer Jump $-$                                                                       \\
 3646.50 &   0.11& $\pm$   0.04&  0.15& $\pm$   0.06 & Balmer Jump $+$                                                                       \\
 3705.02 &   0.85& $\pm$   0.16&  1.12& $\pm$   0.21 &  He~{\sc i}      &  V25       &  2p 3P*    &  7d 3D     &          9 &       15    \\
 3711.97 &   0.92& $\pm$   0.14&  1.21& $\pm$   0.19 &  H~{\sc i}       &  H15       &  2p+ 2P*   &  15d+ 2D   &          8 &        *    \\
 3713.08 &   0.61& $\pm$   0.15&  0.80& $\pm$   0.19 &  Ne~{\sc ii}     &  V5        &  3s 2P     &  3p 2D*    &          4 &        6    \\
 3726.03 &  30.21& $\pm$   3.25& 39.72& $\pm$   4.33 &  [O~{\sc ii}]    &  F1        &  2p3 4S*   &  2p3 2D*   &          4 &        4    \\
 3728.82 &  25.06& $\pm$   3.26& 32.96& $\pm$   4.32 &  [O~{\sc ii}]    &  F1        &  2p3 4S*   &  2p3 2D*   &          4 &        6    \\
 3750.15 &   2.45& $\pm$   0.09&  3.21& $\pm$   0.13 &  H~{\sc i}       &  H12       &  2p+ 2P*   &  12d+ 2D   &          8 &        *    \\
 3759.87 &   0.46& $\pm$   0.07&  0.61& $\pm$   0.09 &  O~{\sc iii}     &  V2        &  3s 3P*    &  3p 3D     &          5 &        7    \\
 3770.63 &   3.03& $\pm$   0.09&  3.95& $\pm$   0.14 &  H~{\sc i}       &  H11       &  2p+ 2P*   &  11d+ 2D   &          8 &        *    \\
 3797.90 &   3.76& $\pm$   0.08&  4.88& $\pm$   0.13 &  H~{\sc i}       &  H10       &  2p+ 2P*   &  10d+ 2D   &          8 &        *    \\
 3819.62 &   1.52& $\pm$   0.07&  1.97& $\pm$   0.09 &  He~{\sc i}      &  V22       &  2p 3P*    &  6d 3D     &          9 &       15    \\
 3834.89 &   5.16& $\pm$   0.24&  6.64& $\pm$   0.33 &  He~{\sc ii}     &  4.18      &  4f+ 2F*   &  18g+ 2G   &         32 &        *    \\
 3835.39 &  *     &             & *     &              &  H~{\sc i}       &  H9        &  2p+ 2P*   &  9d+ 2D    &          8 &        *          \\
 3868.75 &  44.68& $\pm$   0.57& 57.15& $\pm$   1.15 &  [Ne~{\sc iii}]  &  F1        &  2p4 3P    &  2p4 1D    &          5 &        5    \\
 3888.65 &  19.06& $\pm$   0.24& 24.28& $\pm$   0.49 &  He~{\sc i}      &  V2        &  2s 3S     &  3p 3P*    &          3 &        9    \\
 3889.05 &  *     &             & *     &              &  H~{\sc i}       &  H8        &  2p+ 2P*   &  8d+ 2D    &          8 &        *          \\
 3926.54 &   0.20& $\pm$   0.03&  0.26& $\pm$   0.04 &  He~{\sc i}      &  V58       &  2p 1P*    &  8d 1D     &          3 &        5    \\
 3964.73 &   0.96& $\pm$   0.19&  1.21& $\pm$   0.24 &  He~{\sc i}      &  V5        &  2s 1S     &  4p 1P*    &          1 &        3    \\
 3967.46 &   9.24& $\pm$   0.41& 11.58& $\pm$   0.54 &  [Ne~{\sc iii}]  &  F1        &  2p4 3P    &  2p4 1D    &          3 &        5    \\
 3970.07 &  11.79& $\pm$   0.41& 14.75& $\pm$   0.56 &  H~{\sc i}       &  H7        &  2p+ 2P*   &  7d+ 2D    &          8 &       98          \\
 4009.26 &   0.29& $\pm$   0.03&  0.37& $\pm$   0.04 &  He~{\sc i}      &  V55       &  2p 1P*    &  7d 1D     &          3 &        5    \\
 4026.08 &   3.05& $\pm$   0.11&  3.77& $\pm$   0.14 &  N~{\sc ii}      &  V39b      &  3d 3F*    &  4f 2[5]   &          7 &        9    \\
 4026.21 &  *     &             & *     &              &  He~{\sc i}      &  V18       &  2p 3P*    &  5d 3D     &          9 &       15    \\
 4041.31 &   0.44& $\pm$   0.06&  0.54& $\pm$   0.07 &  N~{\sc ii}      &  V39b      &  3d 3F*    &  4f 2[5]   &          9 &       11    \\
 4068.60 &   1.51& $\pm$   0.10&  1.84& $\pm$   0.12 &  [S~{\sc ii}]    &  F1        &  2p3 4S*   &  2p3 2P*   &          4 &        4    \\
 4069.62 &   0.98& $\pm$   0.10&  1.19& $\pm$   0.12 &  O~{\sc ii}      &  V10       &  3p 4D*    &  3d 4F     &          2 &        4    \\
 4069.89 &  *     &             & *     &              &  O~{\sc ii}      &  V10       &  3p 4D*    &  3d 4F     &          4 &        6    \\
 4072.16 &   0.97& $\pm$   0.10&  1.19& $\pm$   0.12 &  O~{\sc ii}      &  V10       &  3p 4D*    &  3d 4F     &          6 &        8    \\
 4075.86 &   1.59& $\pm$   0.10&  1.94& $\pm$   0.12 &  O~{\sc ii}      &  V10       &  3p 4D*    &  3d 4F     &          8 &       10    \\
 4076.35 &  *     &             & *     &              &  [S~{\sc ii}]    &  F1        &  2p3 4S*   &  2p3 2P*   &          2 &        4    \\
 4089.29 &   0.59& $\pm$   0.12&  0.72& $\pm$   0.14 &  O~{\sc ii}      &  V48a      &  3d 4F     &  4f G5*    &         10 &       12    \\
 4097.25 &   1.93& $\pm$   0.63&  2.34& $\pm$   0.77 &  O~{\sc ii}      &  V48b      &  3d 4F     &  4f G4*    &          8 &       10    \\
 4097.26 &  *     &             & *     &              &  O~{\sc ii}      &  V48b      &  3d 4F     &  4f G4*    &          8 &       10    \\
 4097.33 &  *     &             & *     &              &  N~{\sc iii}     &  V1        &  3s 2S     &  3p 2P*    &          2 &        4    \\
 4101.74 &  20.77& $\pm$   1.31& 25.24& $\pm$   1.55 &  H~{\sc i}       &  H6        &  2p+ 2P*   &  6d+ 2D    &          8 &       72          \\
 4119.22 &   0.28& $\pm$   0.04&  0.34& $\pm$   0.05 &  O~{\sc ii}      &  V20       &  3p 4P*    &  3d 4D     &          6 &        8    \\
 4120.28 &   0.28& $\pm$   0.03&  0.33& $\pm$   0.04 &  O~{\sc ii}      &  V20       &  3p 4P*    &  3d 4D     &          6 &        6    \\
 4120.54 &  *     &             & *     &              &  O~{\sc ii}      &  V20       &  3p 4P*    &  3d 4D     &          6 &        4    \\
 4120.84 &  *     &             & *     &              &  He~{\sc i}      &  V16       &  2p 3P*    &  5s 3S     &          9 &        3    \\
 4121.46 &   0.19& $\pm$   0.03&  0.23& $\pm$   0.03 &  O~{\sc ii}      &  V19       &  3p 4P*    &  3d 4P     &          2 &        2    \\
 4132.80 &   0.23& $\pm$   0.02&  0.28& $\pm$   0.03 &  O~{\sc ii}      &  V19       &  3p 4P*    &  3d 4P     &          2 &        4    \\
 4143.76 &   0.43& $\pm$   0.04&  0.52& $\pm$   0.05 &  He~{\sc i}      &  V53       &  2p 1P*    &  6d 1D     &          3 &        5    \\
 4153.30 &   0.38& $\pm$   0.04&  0.46& $\pm$   0.05 &  O~{\sc ii}      &  V19       &  3p 4P*    &  3d 4P     &          4 &        6    \\
 4168.97 &   0.15& $\pm$   0.04&  0.17& $\pm$   0.04 &  He~{\sc i}      &  V52       &  2p 1P*    &  6s 1S     &          3 &        1    \\
 4169.22 &  *     &             & *     &              &  O~{\sc ii}      &  V19       &  3p 4P*    &  3d 4P     &          6 &        6    \\
 4195.76 &   0.11& $\pm$   0.03&  0.13& $\pm$   0.04 &  N~{\sc iii}     &  V6        &  3s' 2P*   &  3p' 2D    &          2 &        4    \\
 4199.83 &   0.20& $\pm$   0.03&  0.24& $\pm$   0.03 &  He~{\sc ii}     &  4.11      &  4f+ 2F*   &  11g+ 2G   &         32 &        *    \\
 4200.10 &  *     &             & *     &              &  N~{\sc iii}     &  V6        &  3s' 2P*   &  3p' 2D    &          4 &        6    \\
 4219.37 &   0.10& $\pm$   0.03&  0.12& $\pm$   0.03 &  Ne~{\sc ii}     &  V52a      &  3d 4D     &  4f 2[4]*  &          8 &        8    \\
 4219.74 &  *     &             & *     &              &  Ne~{\sc ii}     &  V52a      &  3d 4D     &  4f 2[4]*  &          8 &       10    \\
 4236.91 &   0.31& $\pm$   0.03&  0.37& $\pm$   0.04 &  N~{\sc ii}      &  V48a      &  3d 3D*    &  4f 1[3]   &          3 &        5    \\
 4237.05 &  *     &             & *     &              &  N~{\sc ii}      &  V48b      &  3d 3D*    &  4f 1[4]   &          5 &        7    \\
 4241.78 &   0.41& $\pm$   0.02&  0.49& $\pm$   0.02 &  N~{\sc ii}      &  V48b      &  3d 3D*    &  4f 1[4]   &          7 &        9    \\
 4267.15 &   1.59& $\pm$   0.06&  1.86& $\pm$   0.07 &  C~{\sc ii}      &  V6        &  3d 2D     &  4f 2F*    &         10 &       14    \\
 4275.55 &   0.46& $\pm$   0.03&  0.53& $\pm$   0.04 &  O~{\sc ii}      &  V67a      &  3d 4D     &  4f F4*    &          8 &       10    \\
 4275.99 &  *     &             & *     &              &  O~{\sc ii}      &  V67b      &  3d 4D     &  4f F3*    &          4 &        6    \\
 4276.28 &  *     &             & *     &              &  O~{\sc ii}      &  V67b      &  3d 4D     &  4f F3*    &          6 &        6    \\
 4276.75 &  *     &             & *     &              &  O~{\sc ii}      &  V67b      &  3d 4D     &  4f F3*    &          6 &        8    \\
 4277.43 &   0.16& $\pm$   0.03&  0.19& $\pm$   0.04 &  O~{\sc ii}      &  V67c      &  3d 4D     &  4f F2*    &          2 &        4    \\
 4277.89 &  *     &             & *     &              &  O~{\sc ii}      &  V67b      &  3d 4D     &  4f F3*    &          8 &        8    \\
   \hline
\noalign{\smallskip}
\end{tabular}
\end{table*}

\begin{table*}
\contcaption{Observed and dereddened emission line fluxes from NGC~6778}
\begin{tabular}{lrlrlllllll}
\hline
    $ \lambda $ (\AA{}) & \multicolumn{2}{c}{$F \left( \lambda \right) $ }& \multicolumn{2}{c}{$I \left( \lambda \right) $} & Ion & Multiplet & Lower term & Upper term & g$_1$ & g$_2$\\
\hline
 4283.73 &   0.13& $\pm$   0.04&  0.15& $\pm$   0.05 &  O~{\sc ii}      &  V67c      &  3d 4D     &  4f F2*    &          4 &        4    \\
 4291.25 &   0.13& $\pm$   0.03&  0.15& $\pm$   0.04 &  O~{\sc ii}      &  V55       &  3d 4P     &  4f G3*    &          6 &        8    \\
 4294.78 &   0.15& $\pm$   0.03&  0.17& $\pm$   0.03 &  O~{\sc ii}      &  V53b      &  3d 4P     &  4f D2*    &          4 &        6    \\
 4294.92 &  *     &             & *     &              &  O~{\sc ii}      &  V53b      &  3d 4P     &  4f D2*    &          4 &        4    \\
4303.61 &   0.40& $\pm$   0.04&  0.46& $\pm$   0.05 &  O~{\sc ii}      &  V65a      &  3d 4D     &  4f G5*    &          8 &       10    \\
 4303.82 &  *     &             & *     &              &  O~{\sc ii}      &  V53a      &  3d 4P     &  4f D3*    &          6 &        8    \\
 4317.14 &   0.22& $\pm$   0.05&  0.25& $\pm$   0.06 &  O~{\sc ii}      &  V2        &  3s 4P     &  3p 4P*    &          2 &        4    \\
 4340.47 &  41.56& $\pm$   0.37& 47.58& $\pm$   0.55 &  H~{\sc i}       &  H5        &  2p+ 2P*   &  5d+ 2D    &          8 &       50          \\
 4363.21 &   1.82& $\pm$   0.05&  2.07& $\pm$   0.06 &  [O~{\sc iii}]   &  F2        &  2p2 1D    &  2p2 1S    &          5 &        1    \\
 4366.89 &   0.28& $\pm$   0.06&  0.32& $\pm$   0.07 &  O~{\sc ii}     &  V2        &  3s 4P     &  3p 4P*    &          6 &        4    \\
 4379.11 &   0.66& $\pm$   0.04&  0.74& $\pm$   0.05 &  N~{\sc iii}     &  V18       &  4f 2F*    &  5g 2G     &         14 &       18    \\
 4379.55 &  *     &             & *     &              &  Ne~{\sc ii}     &  V60b      &  3d 2F     &  4f 1[4]*  &          8 &       10    \\
 4387.93 &   0.90& $\pm$   0.04&  1.02& $\pm$   0.04 &  He~{\sc i}      &  V51       &  2p 1P*    &  5d 1D     &          3 &        5    \\
 4391.99 &   0.15& $\pm$   0.04&  0.17& $\pm$   0.04 &  Ne~{\sc ii}     &  V55e      &  3d 4F     &  4f 2[5]*  &         10 &       10    \\
 4392.00 &  *     &             & *     &              &  Ne~{\sc ii}     &  V55e      &  3d 4F     &  4f 2[5]*  &         10 &       10    \\
 4409.30 &   0.11& $\pm$   0.02&  0.12& $\pm$   0.02 &  Ne~{\sc ii}     &  V55e      &  3d 4F     &  4f 2[5]*  &          8 &       10    \\
 4414.90 &   0.15& $\pm$   0.03&  0.16& $\pm$   0.03 &  O~{\sc ii}      &  V5        &  3s 2P     &  3p 2D*    &          4 &        6    \\
 4416.97 &   0.15& $\pm$   0.03&  0.17& $\pm$   0.03 &  O~{\sc ii}      &  V5        &  3s 2P     &  3p 2D*    &          2 &        4    \\
 4432.74 &   0.22& $\pm$   0.05&  0.25& $\pm$   0.05 &  N~{\sc ii}      &  V55b      &  3d 3P*    &  4f 2[3]   &          5 &        7    \\
 4432.75 &  *     &             & *     &              &  N~{\sc ii}      &  V55b      &  3d 3P*    &  4f 2[3]   &          5 &        7    \\
 4471.50 &   6.94& $\pm$   0.11&  7.69& $\pm$   0.13 &  He~{\sc i}      &  V14       &  2p 3P*    &  4d 3D     &          9 &       15    \\
 4491.07 &   0.19& $\pm$   0.03&  0.21& $\pm$   0.03 &  C~{\sc ii}      &            &  4f 2F*    &  9g 2G     &         14 &       18    \\
 4491.23 &  *     &             & *     &              &  O~{\sc ii}      &  V86a      &  3d 2P     &  4f D3*    &          4 &        6    \\
 4510.91 &   0.15& $\pm$   0.03&  0.16& $\pm$   0.03 &  N~{\sc iii}     &  V3        &  3s' 4P*   &  3p' 4D    &          2 &        4    \\
 4530.41 &   0.23& $\pm$   0.02&  0.25& $\pm$   0.02 &  N~{\sc ii}      &  V58b      &  3d 1F*    &  4f 2[5]   &          7 &        9    \\
 4530.86 &  *     &             & *     &              &  N~{\sc iii}     &  V3        &  3s' 4P*   &  3p' 4D    &          4 &        2    \\
 4534.58 &   0.06& $\pm$   0.01&  0.06& $\pm$   0.02 &  N~{\sc iii}     &  V3        &  3s' 4P*   &  3p' 4D    &          6 &        6    \\
 4541.59 &   0.22& $\pm$   0.03&  0.24& $\pm$   0.03 &  He~{\sc ii}     &  4.9       &  4f+ 2F*   &  9g+ 2G    &         32 &        *    \\
 4552.53 &   0.15& $\pm$   0.03&  0.16& $\pm$   0.03 &  N~{\sc ii}      &  V58a      &  3d 1F*    &  4f 2[4]   &          7 &        9    \\
 4562.60 &   0.14& $\pm$   0.02&  0.15& $\pm$   0.02 &  Mg~{\sc i}]     &            &  3s2 1S    &  3s3p 3P*  &          1 &        5    \\
 4571.10 &   0.18& $\pm$   0.02&  0.19& $\pm$   0.02 &  Mg~{\sc i}]     &            &  3s2 1S    &  3s3p 3P*  &          1 &        3    \\
 4602.13 &   0.17& $\pm$   0.03&  0.18& $\pm$   0.04 &  O~{\sc ii}      &  V92b      &  3d 2D     &  4f F3*    &          4 &        6    \\
 4607.03 &   0.11& $\pm$   0.03&  0.11& $\pm$   0.03 &  [Fe~{\sc iii}]  &  F3        &  3d6 5D    &  3d6 3F2   &          9 &        7    \\
 4607.16 &  *     &             & *     &              &  N~{\sc ii}      &  V5        &  3s 3P*    &  3p 3P     &          1 &        3    \\
 4609.44 &   0.30& $\pm$   0.04&  0.33& $\pm$   0.04 &  O~{\sc ii}      &  V92a      &  3d 2D     &  4f F4*    &          6 &        8    \\
 4621.25 &   0.13& $\pm$   0.03&  0.14& $\pm$   0.04 &  O~{\sc ii}      &  V92       &  3d 2D     &  4f 2[2]*  &          6 &        6    \\
 4621.39 &  *     &             & *     &              &  N~{\sc ii}      &  V5        &  3s 3P*    &  3p 3P     &          3 &        1    \\
 4630.54 &   0.31& $\pm$   0.05&  0.33& $\pm$   0.05 &  N~{\sc ii}      &  V5        &  3s 3P*    &  3p 3P     &          5 &        5    \\
 4634.14 &   0.74& $\pm$   0.06&  0.78& $\pm$   0.06 &  N~{\sc iii}     &  V2        &  3p 2P*    &  3d 2D     &          2 &        4    \\
 4638.86 &   0.55& $\pm$   0.08&  0.58& $\pm$   0.08 &  O~{\sc ii}      &  V1        &  3s 4P     &  3p 4D*    &          2 &        4    \\
 4640.64 &   1.70& $\pm$   0.07&  1.81& $\pm$   0.08 &  N~{\sc iii}     &  V2        &  3p 2P*    &  3d 2D     &          4 &        6    \\
 4641.81 &   1.11& $\pm$   0.07&  1.18& $\pm$   0.07 &  O~{\sc ii}      &  V1        &  3s 4P     &  3p 4D*    &          4 &        6    \\
 4641.84 &  *     &             & *     &              &  N~{\sc iii}     &  V2        &  3p 2P*    &  3d 2D     &          4 &        4    \\
 4649.13 &   1.63& $\pm$   0.06&  1.72& $\pm$   0.07 &  O~{\sc ii}      &  V1        &  3s 4P     &  3p 4D*    &          6 &        8    \\
 4650.84 &   0.44& $\pm$   0.06&  0.47& $\pm$   0.07 &  O~{\sc ii}      &  V1        &  3s 4P     &  3p 4D*    &          2 &        2    \\
 4661.63 &   0.54& $\pm$   0.03&  0.57& $\pm$   0.03 &  O~{\sc ii}      &  V1        &  3s 4P     &  3p 4D*    &          4 &        4    \\
 4673.73 &   0.09& $\pm$   0.03&  0.10& $\pm$   0.03 &  O~{\sc ii}      &  V1        &  3s 4P     &  3p 4D*    &          4 &        2    \\
 4676.24 &   0.41& $\pm$   0.03&  0.43& $\pm$   0.03 &  O~{\sc ii}      &  V1        &  3s 4P     &  3p 4D*    &          6 &        6    \\
 4685.68 &   6.21& $\pm$   0.51&  6.50& $\pm$   0.52 &  He~{\sc ii}     &  3.4       &  3d+ 2D    &  4f+ 2F*   &         18 &       32    \\
 4696.35 &   0.10& $\pm$   0.03&  0.10& $\pm$   0.03 &  O~{\sc ii}      &  V1        &  3s 4P     &  3p 4D*    &          6 &        4    \\
 4711.37 &   0.84& $\pm$   0.05&  0.88& $\pm$   0.05 &  [Ar~{\sc iv}]   &  F1        &  3p3 4S*   &  3p3 2D*   &          4 &        6    \\
 4713.17 &   0.68& $\pm$   0.05&  0.70& $\pm$   0.05 &  He~{\sc i}      &  V12       &  2p 3P*    &  4s 3S     &          9 &        3    \\
 4740.17 &   0.70& $\pm$   0.05&  0.73& $\pm$   0.05 &  [Ar~{\sc iv}]   &  F1        &  3p3 4S*   &  3p3 2D*   &          4 &        4    \\
 4788.13 &   0.08& $\pm$   0.02&  0.08& $\pm$   0.02 &  N~{\sc ii}      &  V20       &  3p 3D     &  3d 3D*    &          5 &        5    \\
 4803.29 &   0.18& $\pm$   0.01&  0.18& $\pm$   0.01 &  N~{\sc ii}      &  V20       &  3p 3D     &  3d 3D*    &          7 &        7    \\
 4815.55 &   0.05& $\pm$   0.01&  0.05& $\pm$   0.01 &  S~{\sc ii}      &  V9        &  4s 4P     &  4p 4S*    &          6 &        4    \\
 4861.33 & 100.00& $\pm$   2.15&100.00& $\pm$   2.15 &  H~{\sc i}       &  H4        &  2p+ 2P*   &  4d+ 2D    &          8 &       32          \\
 4906.83 &   0.12& $\pm$   0.02&  0.12& $\pm$   0.02 &  [Fe~{\sc iv}]   &  V28       &  3p 4S*    &  3d 4P     &          4 &        4    \\
 4921.93 &   2.15& $\pm$   0.04&  2.12& $\pm$   0.04 &  He~{\sc i}      &  V48       &  2p 1P*    &  4d 1D     &          3 &        5    \\
 4924.53 &   0.24& $\pm$   0.04&  0.24& $\pm$   0.04 &  O~{\sc ii}      &  V28       &  3p 4S*    &  3d 4P     &          4 &        6    \\
 4958.91 & 175.50& $\pm$   3.01&171.11& $\pm$   2.97 &  [O~{\sc iii}]   &  F1        &  2p2 3P    &  2p2 1D    &          3 &        5    \\
 5875.66 &  27.81& $\pm$   0.19& 22.12& $\pm$   0.35 &  He~{\sc i}      &  V11       &  2p 3P*    &  3d 3D     &          9 &       15    \\
 5931.78 &   0.10& $\pm$   0.03&  0.08& $\pm$   0.02 &  N~{\sc ii}      &  V28       &  3p 3P     &  3d 3D*    &          3 &        5    \\
 5941.65 &   0.26& $\pm$   0.03&  0.20& $\pm$   0.02 &  N~{\sc ii}      &  V28       &  3p 3P     &  3d 3D*    &          5 &        7    \\
    \hline
\noalign{\smallskip}
\end{tabular}
\end{table*}

\begin{table*}
\contcaption{Observed and dereddened emission line fluxes from NGC~6778}
\begin{tabular}{lrlrlllllll}
\hline
    $ \lambda $ (\AA{}) & \multicolumn{2}{c}{$F \left( \lambda \right) $ }& \multicolumn{2}{c}{$I \left( \lambda \right) $} & Ion & Multiplet & Lower term & Upper term & g$_1$ & g$_2$\\
\hline
 6300.34 &   4.98& $\pm$   0.05&  3.69& $\pm$   0.08 &  [O~{\sc i}]     &  F1        &  2p4 3P    &  2p4 1D    &          5 &        5    \\
 6312.10 &   1.42& $\pm$   0.03&  1.05& $\pm$   0.03 &  [S~{\sc iii}]   &  F3        &  2p2 1D    &  2p2 1S    &          5 &        1    \\
 6347.10 &   0.12& $\pm$   0.02&  0.09& $\pm$   0.02 &  Si~{\sc ii}     &  V2        &  4s 2S     &  4p 2P*    &          2 &        4    \\
 6363.78 &   1.66& $\pm$   0.03&  1.22& $\pm$   0.03 &  [O~{\sc i}]     &  F1        &  2p4 3P    &  2p4 1D    &          3 &        5    \\
 6371.38 &   0.18& $\pm$   0.03&  0.13& $\pm$   0.03 &  S~{\sc iii}     &  V2        &  4s 2S     &  4p 2P*    &          2 &        2    \\
 6461.95 &   0.23& $\pm$   0.04&  0.17& $\pm$   0.03 &  C~{\sc ii}      &            &  4f 2F*    &  6g 2G     &         14 &       18    \\
 6548.10 & 104.35& $\pm$   2.69& 74.37& $\pm$   2.47 &  [N~{\sc ii}]    &  F1        &  2p2 3P    &  2p2 1D    &          3 &        5    \\
 6562.77 & 407.94& $\pm$   7.66&289.88& $\pm$   3.91 &  H~{\sc i}       &  H3        &  2p+ 2P*   &  3d+ 2D    &          8 &       18          \\
 6583.50 & 319.79& $\pm$   6.03&226.65& $\pm$   6.50 &  [N~{\sc ii}]    &  F1        &  2p2 3P    &  2p2 1D    &          5 &        5    \\
 6678.16 &   9.46& $\pm$   0.07&  6.61& $\pm$   0.16 &  He~{\sc i}      &  V46       &  2p 1P*    &  3d 1D     &          3 &        5    \\
 6716.44 &  24.35& $\pm$   0.22& 16.92& $\pm$   0.41 &  [S~{\sc ii}]    &  F2        &  2p3 4S*   &  2p3 2D*   &          4 &        6    \\
 6730.82 &  25.96& $\pm$   0.23& 18.00& $\pm$   0.44 &  [S~{\sc ii}]    &  F2        &  2p3 4S*   &  2p3 2D*   &          4 &        4    \\
 6795.00 &   0.06& $\pm$   0.02&  0.04& $\pm$   0.01 &  [K~{\sc iv}]    &  F1        &  3p4 3P    &  3p4 1D    &          3 &        5    \\
 7065.25 &   6.01& $\pm$   0.23&  3.98& $\pm$   0.18 &  He~{\sc i}      &  V10       &  2p 3P*    &  3s 3S     &          9 &        3    \\
\hline
\end{tabular}
\end{table*}

\subsection{Influence of atomic data on results}

The atomic data included in \textsc{neat} is summarised in \citet{wesson12}.  We used the default data for all ions with the exception of [S~{\sc ii}], for which we found that the default data set \citep[collision strengths and Einstein \textit{A}-values from the CHIANTI database;][]{landi12} resulted in a temperature significantly lower than that derived from [O~\textsc{iii}].  We therefore performed the analysis using atomic data from multiple sources in order to investigate their influence on the final results.  We used all combinations of A-values and collision strengths from \citet{ramsbottom96}, \citet{tayal10}, \citet{podobedova09}, and \citet{keenan93}.  The reported [S~\textsc{ii}] density varies by almost a factor of 2 (ranging from roughly 500--900 cm$^{-3}$), and the temperature between approximately 7000--9500K.  For the final analysis, presented in Tables~\ref{tab:etd}, \ref{tab:ionic}, \ref{tab:abund}, and \ref{tab:O2+}, the data from \citet{podobedova09} and \citet{tayal10} were used as those values provided the best agreement with the temperatures derived from [O~\textsc{iii}].  However, the observed variation in determined parameters with atomic data used demonstrates that choice of atomic data is a source of considerable systematic uncertainty, which is additional to the statistical uncertainties reported in our analysis.

\subsection{Rescaling of the red spectrum}
As described in Section \ref{sec:obs}, the fluxes derived from the ``red'' spectrum were scaled such that the reddening calculated from the ratio of H$\alpha$/H$\beta$ matched that derived from the ratios of the Balmer lines observed in the ``blue'' spectrum (the scaling amounting to an increase in flux of 10\%).  The observed difference in reddening calculated may be considered to arise from the small spatial offset between the ``red'' and ``blue'' spectra (and is therefore corrected by the rescaling), or may actually be an intrinsic property of the nebulae (whereby the rescaling would be erroneous).  \citet{bohigas15} states that an observed difference in the reddening calculated by different ratios of Balmer lines can naturally arise from the presence of a cool component in the nebula (as the ratios all have varying dependencies on temperature), with the H$\alpha$/H$\beta$ ratio being the one most affected.  However, this effect would result in an H$\alpha$/H$\beta$ extinction greater than that calculated for other Balmer ratios, whereas we find the inverse for our unscaled spectrum.  Furthermore, for this effect to be significant, the ratio of H$^+$ mass in the cold and hot components, $M_c/M_h$, has to be appreciable - which has been shown not to be the case in several other high \emph{adf} PNe with binary central stars \citep{corradi15}.  We, therefore, show all our results for the case where the ``red'' spectrum fluxes have been rescaled (noting that the effect on the results is small in any case).
 
 \subsection{Helium abundance and the temperature of the cold component}
 \label{sec:hetemp}
 
The abundances determined generally agree well with those found in the previous studies (for those species where prior measurements exist), particularly the elevated helium abundance which was suggested by \citet{czyzak73} to be an artefact of direct line excitation.  This explanation seems unlikely given that there is no such evidence presented by the C~\textsc{ii} and O~\textsc{ii} lines, and that the central star does not appear to be so luminous \citep{guerrero12}.  \textsc{neat} derives the total helium abundance from the ionic abundances of the 4471, 5876 and 6678~\AA{} He~\textsc{i} lines (weighted 1:3:1) and using the adopted temperature, $T_e=8800$ K.  The agreement between the abundances from each line at this temperature is rather poor, with values of 0.148, 0.155 and 0.163 respectively.  From the moderate temperature dependence of the $\lambda$6678/4471 and $\lambda$5876/4471 line ratios, we derive much lower temperatures of $T_e\sim$4300 K for 5876/4471, and $T_e\sim$3400 K for 6678/4471.  In addition, the individual line abundances agree much more closely at these lower temperatures - for the average temperature derived from these ratios, $T_e=3950$ K, the lines give abundances of 0.134, 0.132 and 0.135 respectively.  Still better agreement is found at even lower temperatures - we calculated the helium abundances at temperatures from 2\,000--10\,000\,K, and found that the lowest standard deviation of the three values occurred at $\sim$3\,000\,K.

The lower temperature determined from the He~\textsc{i} line ratios is widely observed in planetary nebulae \citep[e.g.\ ][]{liu03,tsamis04,wesson05}, and is consistent with a significant fraction of the helium emission arising from within cool hydrogen-deficient clumps \citep{liu06}.  The helium temperatures we derive are very similar to the temperature determined from the Balmer Jump.  Furthermore, the temperature could also be calculated from the ratio of O~\textsc{ii} ORLs following the diagnostics of \citet{mcnabb13}, resulting in a temperature of 1\,300$^{+1\,900}_{-900}$\,K\footnote{The density was also calculated from the O~\textsc{ii} ORL ratios resulting in a value roughly a factor of four greater than those calculated from CELs, but with an extremely large uncertainty.}, even lower than the temperatures calculated from the Balmer Jump and He~\textsc{i} line ratios (but more-or-less agreeing within the uncertainties).

\begin{table}
\centering
\caption{Extinction, temperatures, and densities as determined by the empirical analysis using the \textsc{neat} code.}              
\label{tab:etd}      
\centering                                      
\begin{tabular}{r r l}          
\hline
$c$(H$_\beta$)                &${  0.46}$&$\pm{0.03}$ \\
$T_e$([O~\textsc{iii}]) (K) & $8800$&$\pm{80}$\\
$T_e$([S~\textsc{ii}]) (K) & 8850&$\pm{870}$\\
$T_e$(O~\textsc{ii}$_{\mathrm{ (ORLs)}}$) (K) & 1300 &$^{+1900}_{-900}$\\
$T$(Balmer Jump) (K) & ${ 4090}$&$^{+760}_{ -810}$\\
$n_e$([O~\textsc{ii}]) (cm$^{-3}$)& ${  650}$&$^{+  330}_{ -650}$\\
$n_e$([S~\textsc{ii}]) (cm$^{-3}$)& ${  590}$&$\pm{40}$\\
$n_e$(O~\textsc{ii}$_{\mathrm{ (ORLs)}}$) (cm$^{-3}$)& ${  2000}$&$^{+  8000}_{ -1400}$\\
$n_e$ (adopted) (cm$^{-3}$) & $640$&$^{+  160}_{ -640}$\\
\hline
\end{tabular}
%
\end{table}

\subsection{Calculation of O$^{2+}$ abundance}
As can be seen in Figure~\ref{fig:alfafit}, numerous O~{\sc ii} recombination lines are well detected in our observations.  We derive an O$^{2+}$ abundance by first deriving an abundance for each multiplet from the co-added intensity of the detected lines (assuming $T_e$([O~\textsc{iii}]) as the temperature), and then averaging the abundances from all multiplets.  We excluded some weak and blended lines from the V12 and V20 multiplets, which gave abundances several times higher than other values.  Table \ref{tab:O2+} lists the fractional ionic abundances for each individual line as well as the averages for each multiplet, showing the generally good agreement both within each multiplet and between multiplets
.

 \subsection{Abundance discrepancy factor and spatial variations}
 
  The overall \emph{adf}(O$^{2+}$)\footnote{The \emph{adf}(O$^{2+}$) is defined as the ratio of the total ionic abundance of O$^{2+}$ derived from ORLs to that derived from CELs, i.e. $\frac{\mathrm{O}^{2+}_{\mathrm{ORLs}}}{H^+}$:$\frac{\mathrm{O}^{2+}_{\mathrm{CELs}}}{H^+}$ \citep{liu06b}.} is found to be 17.9, meaning NGC~6778 has one of the highest known PN \emph{adfs}.  As is commonly found in such high \emph{adf} nebulae \citep{liu03,liu06b,tsamis04,liu06}, the \emph{adf} from Ne$^{2+}$ is also found to be in good agreement with that from O$^{2+}$, with a value of 18.0.  The N \emph{adf} is in poorer agreement at 10.7, however this value has a rather large uncertainty as the comparison depends entirely on the assumed ICF - from CELs only the N$^+$ abundance can be directly measured while from the ORLS only N$^{2+}$ can be measured, and in our case the assumed ICF suggests that only $\sim$15\% of N is in the form of N$^+$ so the correction for the CELs is highly uncertain.
  
     \begin{figure}
\centering
\includegraphics[width=\columnwidth]{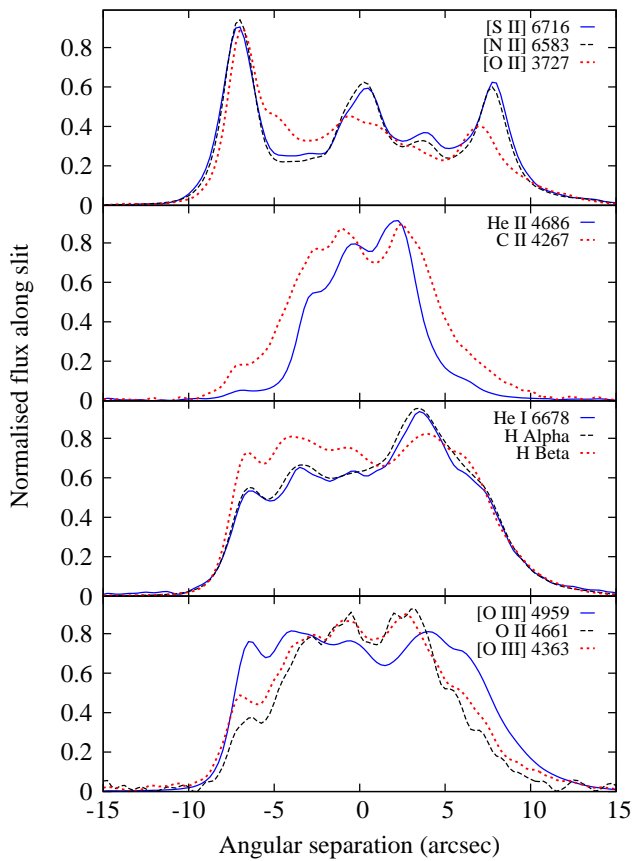}
\caption{Smoothed spatial variations of selected CELs and ORLs along the VLT-FORS2 longslit.  Profiles have been arbitrarily scaled in order to show multiple lines on the same plot, and the x-axis zero corresponds to the part of the slit closest to the geometric centre of the nebula.}
\label{fig:spatial}
\end{figure}
  
Inspecting the spatial variations of the line profiles of CELs and ORLs along the slit provides indications that the abundances are not constant throughout the nebula.   Smoothed spatial profiles, showing the flux variation along the slit of a selection of key CELs and ORLs, are presented in Fig.~\ref{fig:spatial}.  The figure clearly shows that the bright CELs  of [O~\textsc{iii}] and the bright ORLs of  H~\textsc{i}, He~\textsc{i} tend to show a similar distribution, while other ORLs (e.g.\ C~\textsc{ii}~4267\AA{}, O~\textsc{ii}~4649\AA{}) are confined to more central regions.  The profile of H$\alpha$ and H$\beta$ differ slightly; this is almost certainly attributable to the spatial offset between the red and blue exposures - the fact that the two profiles are still rather similar (and the relatively small correction required to match their total flux ratios to the predicted value) clearly shows that this effect is fairly small, and cannot be responsible for the overall differences found between the CELs and ORLs (in any case, most of these are from the blue exposure).
 More intriguingly, the distributions of [S~\textsc{ii}], [N~\textsc{ii}] and [O~\textsc{ii}] appear distinct from those of the other CELs as well those of the ORLs, showing strong, sharp peaks at the nebular centre and extremes.  This is probably due to a combination of ionisation stratification and density variations throughout the nebula.
  
\begin{figure}
\centering
\includegraphics[angle=270,width=\columnwidth]{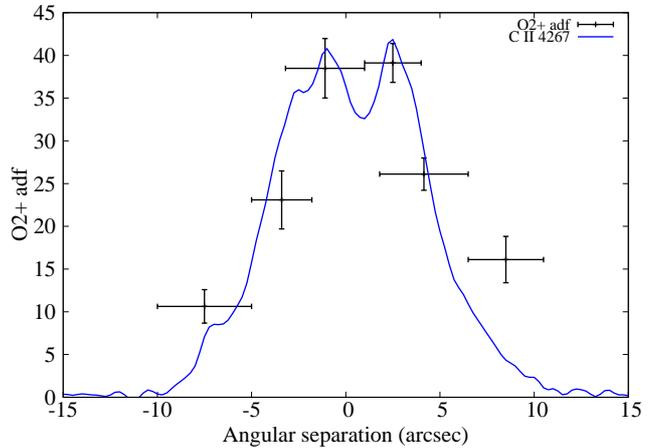}
\caption{The variation of the O$^{2+}$ \emph{adf} along the slit with the profile of the C~\textsc{ii}~4267\AA{} ORL (scaled arbitrarily) which shows a brightness distribution mirroring the \emph{adf}.}
\label{fig:spatial_adf}
\end{figure}

\begin{figure}
\centering
\includegraphics[angle=270,width=\columnwidth]{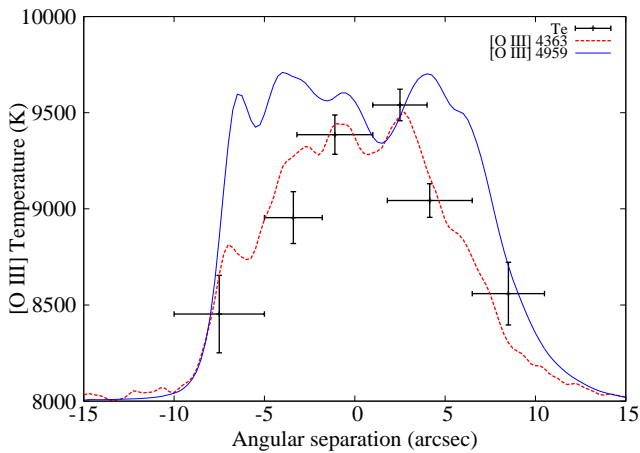}
\caption{The variation of the [O~\textsc{iii}] temperature, $T_e$([O~\textsc{iii}]), along the slit along with the profiles of selected lines of [O~\textsc{iii}] (zeroed and scaled arbitrarily).}
\label{fig:spatial_temp}
\end{figure}

  The differing spatial profiles between CELs and ORLs mean that the \emph{adf} also varies throughout the nebula, reaching above 40 in the very centre and dropping as low as a few in the outer regions.  Similar variations have also been found in other PNe with binary central stars \citep{corradi15}.  To demonstrate this variation, six regions (allowing for good signal-to-noise in each extracted region) along the slit were extracted and the resulting spectra processed using \textsc{alfa} and \textsc{neat}, just as for the summed spectrum.  The resulting \emph{adfs} are displayed in Figure \ref{fig:spatial_adf} over-plotted on the arbitrarily scaled spatial profile of the C~\textsc{ii}~4267\AA{} ORL, showing clearly that the observed \emph{adf} follows the brightness distribution of the ORLs.  Similarly, the temperature is also found to vary across the nebula, following a similar pattern as the \emph{adf} and ORL and [O~\textsc{iii}]~4363\AA{} brightness distributions, as shown in Figure \ref{fig:spatial_temp}.

It is important to note that the spatial profiles of the O~{\sc ii} recombination lines and the [O~\textsc{iii}]~4363\AA{} CEL are remarkably similar, despite the expectation that they should arise from physically distinct, though intermixed, regions of the nebula.  If a substantial amount of oxygen were in the form of O$^{3+}$, then recombination excitation of the $\lambda$4363\AA{} line could contribute a significant fraction of its flux, thus explaining the similarities in line profile.  \citet{liu00} give the following equation for the recombination flux of the $\lambda$4363\AA{} line:

\begin{equation}
\frac{I_R(\lambda4363)}{I(\mathrm{H}\beta)}=12.4t^{0.59}\times \frac{\mathrm{O}^{3+}}{\mathrm{H}^{+}}
\end{equation}

The observed He$^{2+}$/He$^+$ implies that the amount of O in the form of O$^{3+}$ is around 2\%.  For the CEL abundance of 2.80$\times$10$^{-4}$, this gives O$^{3+}$/H$^+$=5.6$\times$10$^{-6}$, and a recombination line $\lambda$4363\AA{} flux of 0.006 where H$\beta$=100.  This is negligible compared to the observed line flux of 2.07$\pm$0.06.  However, adopting the much higher recombination line abundance of 4.73$\times$10$^{-3}$ the recombination line flux would amount to 6.5\% of the observed line flux, sufficient to reduce the calculated [O~{\sc iii}] temperature by $\sim$130\,K, but still much too low to significantly alter the spatial profile of the line.  Thus, there is, as yet, no clear reason for the close similarity between the spatial profiles.

  \begin{table}
\centering
\caption{Ionic abundances, relative to hydrogen, as determined by the empirical analysis using the \textsc{neat} code.  All abundances are calculated using CELs unless otherwise indicated.}              
\label{tab:ionic}      
\centering                                      
\begin{tabular}{r r}          
\hline                        
 Ion & $\frac{X^{i+}}{H^+}$\\    
\hline
He$^{+}$	(ORLs)	     &  *${  0.154}\pm0.003$\\
He$^{2+}$ (ORLs)            & ${  5.27(\pm0.44)\times 10^{ -3}}$\\
C$^{2+}$ (ORLs)   &${  1.73(\pm0.07)\times 10^{ -3}}$\\
N$^{+}$                  &${  5.55(\pm0.15)\times 10^{ -5}}$\\
N$^{2+}$ (ORLs)    &${  3.83(\pm0.16)\times 10^{ -3}}$\\
O$^{+}$                  &${  5.55(\pm0.51)\times 10^{ -5}}$ \\
O$^{2+}$ (CELs)    & ${  2.83(\pm0.15)\times 10^{ -4}}$\\
O$^{2+}$ (ORLs)    &${  5.06(\pm0.23)\times 10^{ -3}}$\\
\emph{adf}(O$^{2+}$)         &17.9\\
Ne$^{2+}$ (CELs)    &${  1.03(\pm0.04)\times 10^{ -4}}$\\
Ne$^{2+}$ (ORLs)   & ${  1.85(\pm0.51)\times 10^{ -3}}$ \\
\emph{adf}(Ne$^{2+}$)         &18.0\\
Ar$^{3+}$                &${  3.80(\pm0.23)\times 10^{ -7}}$\\
S$^{+}$                   &${  1.44(\pm0.07)\times 10^{ -6}}$\\
S$^{2+}$                  &${  4.26(\pm0.26)\times 10^{ -6}}$\\
\hline
\end{tabular}
\medskip

*0.134$\pm$0.01 and 0.128$\pm$0.01 for $T_e=3950$ K and $T_e=3000$ K, respectively
\end{table}

\begin{table}
\centering
\caption{Total nebular abundances, relative to H, of NGC~6778, determined using the \textsc{neat} code (using CELs unless otherwise stated).}              
\label{tab:abund}      
\centering                                      
\begin{tabular}{lr}
\hline
Element & $\frac{X}{H}$\\
\hline
He (ORLs)  &*${  0.159}\pm0.02$\\
C (ORLs) & ${  3.34(\pm0.19)\times 10^{ -3}}$ \\
N   & ${ 4.04 (\pm0.53)\times 10^{ -4}}$\\
N (ORLs) & ${4.33(\pm0.23)\times 10^{-3}}$ \\
\emph{adf}(N) & 10.7\\
O   & ${  3.39(\pm0.16)\times 10^{ -4}}$\\
O (ORLs) & ${  6.06(\pm0.27)\times 10^{ -3}}$\\
\emph{adf}(O) & 17.9\\
Ne & ${  1.36(\pm0.06)\times 10^{ -4}}$\\
Ne (ORLs) & ${  2.22(\pm0.56)\times 10^{ -3}}$\\
\emph{adf}(Ne) & 16.3\\
Ar  &${  4.68(\pm0.31)\times 10^{ -7}}$\\
S   &${  3.39(\pm0.49)\times 10^{ -6}}$\\
\hline
\end{tabular}
\medskip

*0.134$\pm$0.01 and 0.128$\pm$0.01 for $T_e=3950$ K and $T_e=3000$ K, respectively
\end{table}

\begin{table}
\centering
\caption{Fractional ionic abundances for O$^{2+}$ lines and the weighted averages for each multiplet}              
\label{tab:O2+}      
\centering                                      
\begin{tabular}{r r l}          
\hline
$\lambda$ (\AA{}) & Multiplet & $\frac{X(\mathrm{line})}{H^+}$\\
\hline
4649.13 & V1 & 3.46 ($\pm$ 0.14) $\times$10$^{-3}$ \\
4650.84 & V1 & 4.60 ($\pm$ 0.64) $\times$10$^{-3}$ \\
4661.63 & V1 & 4.24 ($\pm$ 0.26) $\times$10$^{-3}$ \\
4673.73 & V1 & 4.65 ($\pm$ 1.33) $\times$10$^{-3}$ \\
4696.35 & V1 & 7.58 ($\pm$ 2.49) $\times$10$^{-3}$ \\
 & {\bf V1} & {\bf 3.89($\pm$ 0.15) $\times$10$^{-3}$}\\
4317.14 & V2 & 3.34 ($\pm$ 0.82) $\times$10$^{-3}$ \\
4366.89 & V2 & 3.79 ($\pm$ 0.89) $\times$10$^{-3}$ \\
 & {\bf V2} & {\bf 3.66($\pm$ 0.61) $\times$10$^{-3}$}\\
4414.90 & V5 & 3.04 ($\pm$ 0.61) $\times$10$^{-3}$ \\
4416.97 & V5 & 5.35 ($\pm$ 1.14) $\times$10$^{-3}$ \\
 & {\bf V5} & {\bf 3.88($\pm$ 0.56) $\times$10$^{-3}$}\\
4072.16 & V10 & 4.80 ($\pm$ 0.57) $\times$10$^{-3}$ \\
 & {\bf V10} & {\bf 4.79($\pm$ 0.57) $\times$10$^{-3}$}\\
4121.46 & V19 & 8.34 ($\pm$ 1.10) $\times$10$^{-3}$ \\
4132.80 & V19 & 5.18 ($\pm$ 0.53) $\times$10$^{-3}$ \\
4153.30 & V19 & 5.64 ($\pm$ 0.81) $\times$10$^{-3}$ \\
 & {\bf V19} & {\bf 5.97($\pm$ 0.44) $\times$10$^{-3}$}\\
4906.83 & V28 & 4.48 ($\pm$ 0.83) $\times$10$^{-3}$ \\
4924.53 & V28 & 5.76 ($\pm$ 0.94) $\times$10$^{-3}$ \\
 & {\bf V28} & {\bf 5.10($\pm$ 0.67) $\times$10$^{-3}$}\\
4089.29 & V48a & 6.19 ($\pm$ 1.30) $\times$10$^{-3}$ \\
4602.13 & V92b & 9.42 ($\pm$ 1.91) $\times$10$^{-3}$ \\
4609.44 & V92a & 6.52 ($\pm$ 0.89) $\times$10$^{-3}$ \\
 & {\bf 3d-4f} & {\bf 7.47($\pm$ 0.84) $\times$10$^{-3}$}\\
 & {\bf O$^{2+}$/H${^+}$} & {\bf 5.06($\pm$ 0.22) $\times$10$^{-3}$}\\
\hline
\end{tabular}
%
\end{table}

\section{Discussion}
\label{sec:discussion}

NGC 6778 presents remarkably strong recombination lines, and the abundances derived from these exceed those from collisionally excited lines by a factor of nearly 20 (rising to around 40 in the centre), making it one of the largest known (see table \ref{tab:adfs}).  High \emph{adf}s have been suggested to indicate the presence of a second nebular component - in addition to the ``standard'' $\sim$$10\,000$K phase with relatively typical chemical composition - which is of a higher metallicity and much lower temperature \citep[too low to excite significant emission from optical CELs; ][]{liu06b,liu06}.

The general abundance pattern (as determined from CELs) of NGC~6778 is not out of the ordinary for a PN, other than an extremely high helium abundance.  It had previously been suggested that this was an effect of direct excitation, but it is more likely due to the presence of low temperature, hydrogen-deficient clumps in the nebula (similarly consistent with its high \emph{adf}).  Accounting for the lower temperature of these clumps ($\sim$$3\,000-4\,000$ K, estimated from the Balmer Jump and He~\textsc{i} temperatures) reduces the derived helium abundance to a more acceptable, though still quite high, level.  The generally high helium abundance may be considered evidence of a massive progenitor, although the non-elevated nitrogen abundance does not seem to support this hypothesis.

\begin{table*}
\caption{The PNe with the highest known \emph{adf}s to-date, as well as the \emph{adf}s of other PNe known to host close-binary central stars.}
\label{tab:adfs}
\begin{tabular}{rrcll}
\hline
PN & \emph{adf} & Period(days) & Notes & References\\
\hline
A~30 & $\gg$100 & N/A & Born-again & \citet{wesson03} \\
A~46 & 120 & 0.47 && \citet{corradi15,afsar08}\\
A~58 & 89 & N/A & Born-again &\citet{wesson08a}\\
Hf~2-2 & 70 &0.40 && \citet{liu06,schaub12}\\
Ou~5 & 56 & 0.36 && \citet{corradi15,corradi14}\\
NGC~1501 & 32 &N/A&&  \citet{ercolano04}\\
M~1-42 & 20 & N/A&& \citet{liu01}\\
NGC~6778 & 18 & 0.15 && This work; \citet{miszalski11b} \\
NGC~40 & 17 & N/A && \citet{liu04}\\
NGC~2022 & 16 & N/A&& \citet{tsamis04} \\
\hline
Hen~2-161 & 11 & $\sim$1 && \citet{jones15a}\\
A~63 & 8 & 0.46 && \citet{corradi15,afsar08}\\
Hen~2-155 & 6 & 0.15 && \citet{jones15a}\\
NGC~5189 & 2 & 4.04 && \citet{garcia-rojas13,manick15}\\
\hline
\end{tabular} 
\medskip

N/A = Not applicable.  Not a confirmed binary.
\end{table*}

It has been highlighted by \citet{corradi15} that the PNe with the highest \emph{adf}s also tend to host close binary central stars.  The high \emph{adf} of NGC~6778 adds further weight to that hypothesis as the nebula is also known to host one of the shortest period binary central stars known \citep{miszalski11b}.  As well as PNe with close-binary central stars, so-called born-again PNe also feature highly on the list of highest known \emph{adfs} \citep{wesson03,wesson08a}.  These are systems where it is thought that, following the ejection of the PN, the central star underwent a very late thermal pulse (VLTP) sometimes known as a final flash, which ejects further, chemically enriched material into the old nebula.  This provides a good explanation for the apparently multiphase nature of these high \emph{adf} nebulae, while the nature of the ``rebirth'' has been strongly linked to binary evolution \citep[adding to the connection between binarity and high \emph{adf}s; ][]{lau11}.  It is important to note that binarity cannot be responsible for the abundance discrepancy problem in general as, for example, binaries are clearly not responsible for the modest \emph{adf}s found in H~\textsc{ii} regions.  This may imply that there are multiple mechanisms which results in \emph{adf}s greater than unity, with binaries playing a key role in forming the most elevated.

Several of the PNe known to show extreme \emph{adf}s display extreme spatial variations in \emph{adf} just as found here in NGC~6778 \citep[Abell~46, for example, shows an \emph{adf}(O$^{2+}$) of over 300 in its central regions; ][]{corradi15}.  The \emph{adf} in NGC~6778 varies across the nebula, and was found to trace well the brightness profiles of the observed ORLs as well as the [O~\textsc{iii}]~4363\AA{} CEL, while other CELs show a different profile.  The \emph{adf} also follows a similar distribution to the temperature as derived from [O~\textsc{iii}] CELs.  Spatial variations, where nebular \emph{adf}s are centrally peaked, could naturally arise from the aforementioned hypothesis whereby freshly processed, metal-enriched material is ejected from the central star after the formation of the main nebular shell, such that the ORL emission is dominated by the central region containing this new material.  There is, however, a strong argument against the VLTP hypothesis in that their abundance patterns do not show the expected carbon enhancement, but appear more akin to Neon Novae \citep{wesson03,wesson08a}.  Novae are an intrinsically binary phenomena, where eruptions occur on the surface of a white dwarf which is accreting material from a companion star.  This connection to novae is supported by the observed properties of the secondaries in those binary central stars subjected to detailed study, where many are found to be ``inflated'' and close to Roche lobe filling \citep[e.g.\ ][]{pollacco94,afsar08}.  For a fuller discussion of the aforementioned hypotheses see \citet{corradi15}.

While all of the very highest PN \emph{adf}s are found in those nebulae with close-binary central stars or those with central stars who are believed to have experienced some form of binary evolution, there are also PNe with binary central stars that do not show such highly elevated \emph{adf}s.  The recent work of \citet{jones15a} found that the PNe Hen~2-155 and Hen~2-161 contained binary central stars and displayed \emph{adf}s of 6 and 11, respectively.  While these values are elevated with respect to the PN average, they are not as extreme as those displayed in the other systems discussed previously (e.g.\ NGC~6778, A~46 and Hf~2-2).  Additionally, NGC~5189 was found to have a particularly low \emph{adf}\footnote{It is interesting to note that the analysis of NGC~5189 where the \emph{adf} was found to be rather small was based on spectra covering a very small nebular volume \citep[mainly focussed on a bright knot; ][]{garcia-rojas13}, and therefore the results may well be somewhat different for the nebula as a whole.} while also playing host to a binary central star \citep{garcia-rojas13,manick15}.  As pointed out by \citet{corradi15}, this means that while a high \emph{adf} is seemingly indicative of a binary evolution, a binary evolution does not neccessarily imply a high \emph{adf}.

The majority of the binary central stars listed in table \ref{tab:adfs} are rather poorly studied, with little more known about them than their orbital period as discovered through photometric or spectroscopic variability.  Comparing their orbital periods to their nebular \emph{adf}s there is no clear pattern,  other than to say that the nebula with the lowest \emph{adf} is also found to host the longest orbital period binary.  However, this is clearly not evidence of a relationship between the two given that the second lowest \emph{adf} is found in the nebula hosting one of the shortest period binaries (a period roughly equal to that of NGC~6778, in fact).  The small numbers and poorly characterised central binaries mean that making any further links between binary parameters and nebular \emph{adf}s is impossible at this time.  It is, however, intriguing to note that the highest \emph{adf} binary, Abell~46, also plays host to one of the lowest mass central stars known -- at $\sim$0.5 M$_\odot$ V477 Lyrae is on the limit between Red Giant Branch (RGB) and Asymptotic Giant Branch (AGB) core masses, possibly symptomatic of an evolutionary link between the high \emph{adf} and central binary.  A low core mass may mean that the central binary passed through the common envelope phase early in its evolution, when its envelope was more bound and therefore more difficult to eject.  As such, the ejected envelope may well be more likely to experience some kind of fallback on to the central star which could prompt an outburst of enriched material, leading to the observed \emph{adf}.  Of the other systems listed in table \ref{tab:adfs}, only Abell~63 and Hen~2-155 have reliable measurements of central star mass, with masses of 0.63 and 0.62 M$_\odot$, respectively (close to the canonical central star mass of 0.6 M$_\odot$).  Until detailed analyses are made of the other binary central stars listed, it will not be possible to explore this possible connection further.

In summary, this work reinforces the link between high \emph{adf}s and central star binarity, and we propose that chemical abundances can be used as an indicator of binarity \citep[just as the presence of jets, rings and axisymmetric structures have previously been used, e.g.\ ][]{boffin12b,jones14a,santander-garcia15}.  We, therefore, strongly encourage the study of other PNe known to show elevated \emph{adf}s in search of binarity, as well as the chemical study of known binary planetary nebulae, in order to probe further their relation.

\section*{Acknowledgments}
We thank the anonymous referee for their comments which improved the detail of this manuscript.

Based on observations made with ESO Telescopes at the La Silla Paranal Observatory under programme ID 093.D-0038.  This work was co-funded under the Marie Curie Actions of the European Commission (FP7-COFUND). This research has been supported by the Spanish Ministry of Economy and Competitiveness (MINECO) under grants AYA2012-35330 and AYA2011-22614.  JGR acknowledges support in the form of a Severo Ochoa excellence program (SEV-2011-0187) postdoctoral fellowship. 

This research has made use of NASA's Astrophysics Data System Bibliographic Services; the SIMBAD database, operated at CDS, Strasbourg, France; the VizieR catalogue access tool, CDS, Strasbourg, France; APLpy, an open-source plotting package for Python hosted at http://aplpy.github.com; Astropy, a community-developed core Python package for Astronomy \citep{astropy}.




\bibliographystyle{mn2e}
\bibliography{literature} 




%
%


\bsp	
\label{lastpage}
\end{document}